\documentclass[10pt,letterpaper]{article}
\usepackage{graphicx}
\usepackage{dcolumn}
\usepackage{bm}
\usepackage{float}
\usepackage{opex3}

\begin{document}

\title{Arrays of open, independently tunable microcavities}

\author{Christian Derntl,$^1$ Michael Schneider,$^2$ Johannes Schalko,$^2$ Achim Bittner,$^2$ J\"{o}rg Schmiedmayer,$^1$ Ulrich Schmid$^2$ and Michael Trupke$^{1,*}$}
\address{$^1$Vienna Center for Quantum Science and Technology, Institute for Atomic and Subatomic Physics, Vienna University of Technology, 1020 Vienna, Austria\\
$^2$Institute for Sensor and Actuator Systems, Vienna University of Technology, 1040 Vienna, Austria}%
\email{$^*$michael.trupke@tuwien.ac.at}

\date{\today}

\begin{abstract}
Optical cavities are of central importance in numerous areas of physics, including precision measurement, cavity optomechanics and cavity quantum electrodynamics. The miniaturisation and scaling to large numbers of sites is of interest for many of these applications, in particular for quantum computation and simulation.
Here we present the first scaled microcavity system which enables the creation of large numbers of highly uniform, tunable light-matter interfaces using ions, neutral atoms or solid-state qubits. The microcavities are created by means of silicon micro-fabrication, are coupled directly to optical fibres and can be independently tuned to the chosen frequency, paving the way for arbitrarily large networks of optical microcavities.
\end{abstract}

 % REPLACE WITH CORRECT OCIS CODES FOR YOUR ARTICLE, MINIMUM OF TWO; Avoid using the OCIS codes for “General” or “General science” whenever possible.

Future technologies based on quantum mechanical effects have the potential to revolutionise our endeavours in metrology, information processing and fundamental physics \cite{ladd,metrology}. Many of the envisaged quantum devices rely on the precise manipulation and readout of large numbers of single quantum systems. Optical microcavities are among the foremost tools in the study and manipulation of the quantum mechanical interaction between light and matter. They have been used to detect single atoms and to generate single photons from quantum emitters including atoms, ions, nitrogen-vacancy (NV) centres in diamond or quantum dots \cite{kimblePhoton, rempePhoton, northupIon, vahala, painter, beausoleil,imamoglu, becher}. Microcavities have furthermore been used to demonstrate the entanglement between atoms and photons, and were instrumental in the creation of an elementary quantum network \cite{kimbleInternet, rempeNetwork}. Large-scale quantum networks, simulators and computers will require vast amounts of physical qubits operating in parallel, as well as optical connections to enable long-distance links between processing sites. Furthermore a large number of proposals have recently been made for the study of the quantum simulation using networks of coupled cavity arrays \cite{hartmannReview, illuminati, lepertIOP}. It is therefore desirable to create a technological platform which enables the enhanced interaction of emitters with light, and their connection to photonic networks.\\
\begin{figure}
\includegraphics[width=\columnwidth]{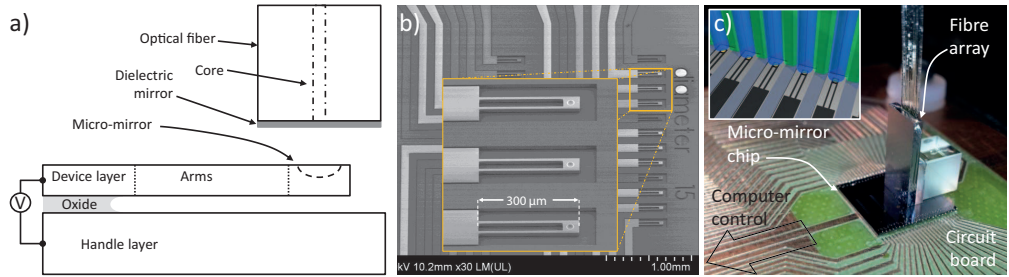}
\caption{\label{fig:combinedFigure1}(a) Schematic of a single microcavity with actuated micro-mirror (not to scale). b) Scanning electron microscope image of a section of the micro-mirror chip.  The inset shows a detailed view of three cantilevers with micro-mirrors. c)  Photograph of the assembled device. Inset: Artist's impression of the assembled microcavity system.  A row of single-mode fibres (blue), each set into a v-groove on a silicon chip (green), is aligned to the equally pitched row of micro-mirrors on cantilevers.}
\end{figure}
 Silicon micro-mirrors have been created by numerous groups and are appealing as microfabrication provides a natural path towards scalability  \cite{trupkeAPL,biedermann,smith}. Fibre-coupled microcavities based on silicon technology have since been used to demonstrate high-fidelity atom detection and photon generation \cite{trupkePRL, goldwin}. More recently, advances in silicon micromachining have led to the improvement of the surface quality of these micro-mirrors and to the achievement of finesse values on the order of $6\times10^4\,\,$   \cite{biedermann, laliotis}. While this finesse is an order of magnitude lower than what has been achieved with macroscopic mirrors, the small mode size in these resonators more than makes up for this by increasing the electric field per photon by more than two orders of magnitude.\\
 Given their direct fabrication in large numbers, the small resulting mode volumes and the high finesse achievable, silicon micro-mirrors offer a promising path to a scalable platform for cavity quantum electrodynamics (CQED). For this purpose, three final requirements need to be fulfilled, namely reproducible coupling of light into and out of each site, single-site tuning to achieve resonance with the emitters, and stable operation.\\
  We show here that all of these requirements can be fulfilled using an approach based on fibre-coupled, silicon-based microcavities, augmented with individual electrostatic tuning for the position of each silicon micro-mirror (see figure \ref{fig:combinedFigure1}).  We demonstrate the fabrication and stable operation of an array of concave microcavity mirrors with independent electrostatic actuation. They are fabricated on doped silicon-on-insulator (SOI) wafers and present mechanical resonance frequencies of over 200 kHz, negligible cross-talk and tuning over more than one free spectral range in the near-infrared. We then describe the operation of a row of twelve fibre-coupled microcavities, and thereby show that this approach can be scaled to arbitrarily large numbers. Finally we outline how these devices will enable the creation of arrays of cavity-enhanced quantum emitters, with numerous applications ranging from information processing to cavity-enhanced sensing and metrology.
\section{Fabrication process and actuation characteristics} We fabricated our structures (see figure \ref{fig:combinedFigure1}) on standard silicon-on-insulator wafers, with a device layer thickness of $50\,\mu$m, a buried oxide layer of $2\,\mu$m thickness and a handle thickness of $400\,\mu$m.  The micro-mirrors are etched into the device layer using an inductively-coupled plasma etching procedure \cite{larsen}, consisting of a masked etch of $70\,$s duration and a maskless etch lasting $7\,$minutes. Next, the entire device is coated with a bi-layer of titanium ($10\,$nm) and gold ($100\,$nm) by e-beam evaporation. This coating acts as a mirror for our interrogation light with $780\,$nm wavelength.  The metallic coating is structured to leave a reflective portion only on the micro-mirrors. The device layer is then structured to reveal the buried oxide layer, using a Bosch process consisting of 100 cycles with a duration of 500 seconds.  This step defines both the cantilevers and the conducting tracks leading to bond pads located at the edges of the chip. The cantilevers are then released by removing the buried oxide with hydrofluoric acid for 15 minutes.  The conducting lines to the bond pads are sufficiently wide to protect part of the oxide they lie on, while the oxide layer is entirely removed from underneath the narrow cantilever arms and the mirror pads. We thereby create a rectangular array of 48 independently addressable cantilevers, consisting of four rows with twelve units. The lateral pitch is $250\,\mu$m, while the distance between rows is $1.75\,$mm.
Each one of our structures (see figure \ref{fig:combinedFigure1}) consists of a square pad with a side length of $L_{p}=50\,\mu$m, supported by two arms of length $L_{A}=250\,\mu$m, width $w_{A}=10\,\mu$m and height $t_{Dev}=22\,\mu$m. We furthermore take into account the under-etch of the supporting device layer at the anchor region of the arms, with a length of $32\,\mu$m and width of $130\,\mu$m. The mirror pad can be actuated by an electrostatic force of magnitude $F_{el}=\epsilon_0 A_{tot} V^2/(2t_{ox}^2)$, where $\epsilon_0$ is the dielectric constant of vacuum, $V$ is the applied voltage, $t_{ox}$ is the oxide thickness and $A_{tot}$ is the combined area of the cantilever pad, the arms and the underetched portion of the anchor region. The composite structure makes an analytical description complex, therefore the device properties were modelled using a finite-element solver (\textsc{Comsol}) to predict resonance frequencies and actuation ranges.
 \begin{figure}
 \includegraphics[width=\columnwidth]{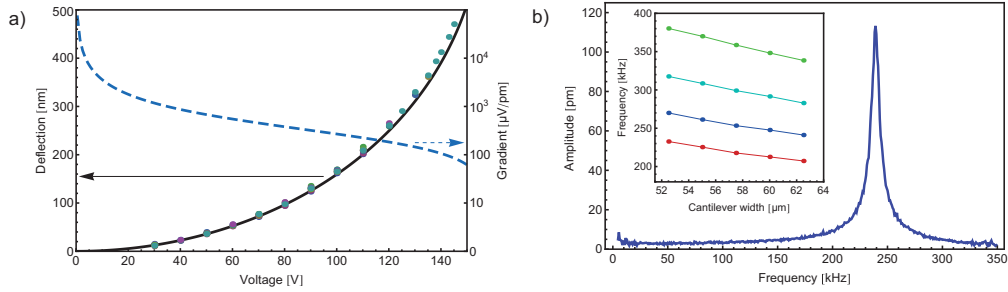}
\caption{\label{fig:figure2}(a) Deflection characteristics of the cantilevers, measured using a white-light interferometer. The solid line shows the deflection as a function of voltage as expected from simulations.  The points show the measured data from 12 cantilevers. The dashed line shows the gradient of the actuation voltage (logarithmic scale, right axis). e) Laser-Doppler vibrometer measurement of the mechanical frequency response of an electrostatically excited cantilever. Inset: Measured mechanical resonance frequencies for cantilever lengths of $225\,\mu$m (green), $250\,\mu$m (cyan), $275\,\mu$m (green) and $300\,\mu$m (red) and varying cantilever width.}
\end{figure}
 Before assembling the device to form the array of microcavities, we characterised the actuation range and resonance frequencies of a variety of cantilevers in order to ascertain their suitability for our application.  Figure \ref{fig:figure2}(a) shows the deflection of 12 cantilevers compared to the simulated performance.  The data indicates high repeatability, as well as good agreement with our expectations.  Furthermore the deflection achieved is sufficient to tune through one free spectral range in the near-infrared, guaranteeing that resonance can be achieved at every site. These measurements demonstrate that the micro-mirrors can be actuated very accurately over most of the accessible deflection range. High-finesse cavities such as those presented in ref \cite{biedermann} require a positional accuracy on the order of $1\,$pm. The actuation gradient for our devices remains larger than $100\,\mu$V/pm throughout most of the actuation range, so that even very high-finesse microcavities can be tuned precisely into a resonance with commercially available voltage supplies. Calculations indicate a pull-in deflection of approximately $667\,$nm for the oxide thickness of 2$\,\mu$m specified by the manufacturer.  Beyond this point, the cantilever will snap downwards and stick irreversibly to the handle layer due to the van der Waals force.  The maximum deflection value can be further increased by using a thicker buried oxide spacer between the device and handle layers, and greater actuation voltages.\\
 The radius of curvature of the micro-mirrors was measured to be $r_C=(56\pm 2)\,\mu$m. It is also possible to create micro-mirrors with radii of curvature of several hundreds of micrometers by varying the mask diameter and etch time.
 In order to lock the microcavities to an optical resonance, a high mechanical resonance frequency far above the acoustic spectrum is desirable. The mechanical spectrum at room temperature of one such cavity is shown in Figure \ref{fig:figure2}(b).  The lowest resonance appears at $239\,$kHz, far beyond the spectrum of acoustic noise sources in a normal laboratory environment. The data in the inset of figure \ref{fig:figure2}(b) show that the cantilever deflection and resonance properties can also be finely tuned throughout a wide range by adjusting the cantilever dimensions.\\
\section{Optical microcavity measurements} For the optical characterisation of the microcavities we place an array of 12 single-mode fibres with a pitch of $250\,\mu$m (manufactured by Ozoptics, Inc.), coated to a reflectivity of $98.7\pm0.1\,\%$ (O.I.B. Jena, GmbH), in front of one row of gold-coated micro-mirrors. We thereby form 12 independently tunable microcavities in a single alignment step (see Appendix). The micro-mirror radius of curvature as well as the cavity length were measured by tuning a free-running diode laser through one free spectral range of a microcavity.  These measurements yielded a curvature of $R_C=(51\pm 1),\mu$m, close to the value obtained in the white-light interferometer measurements.  The cavity length was determined to be $L_C=(42.9\pm 0.5)\,\mu$m, while the cavity half-linewidth $\kappa=2\pi\times(14.97\pm 0.04)\,$GHz.

The fringe visibility $\vartheta=1-P_{min}/P_{max}$ depends on the mode overlap and on the reflectivity values of the fibre mirror and micro-mirror with
\begin{equation}\label{eqn:contrast}
\vartheta=1-\left\vert r_1^a-\frac{r_2 t_1^2 \eta_{FC}^2}{1-r_1 r_2}\right\vert^2,\,\,\, \eta_{FC}=\int_{-\infty}^{\infty}\int_{-\infty}^{\infty}\psi_{F}^{*} \psi_{C}dxdy.
\end{equation}
Here $\psi_F$, $\psi_C$ are the transverse spatial modes of the fibre and cavity, respectively. $r_1$ and $r_2$ are the amplitude reflection coefficients of the fibre and micro-mirror, respectively. $t_1$ is the amplitude transmission coefficient of the fibre mirror, and we include a loss $\ell_1^2=0.1\%$ as specified by the manufacturer (O.I.B. Jena GmbH). Together these coefficients must fulfil the condition $r_1^2+t_1^2+\ell_1^2=1$.
 $\eta_{FC}$ gives the overlap of the fibre and cavity modes. For the lowest-order transverse cavity mode $\eta_{FC}=\frac{2 w_C w_F}{w_C^2+w_F^2}$exp$\left(-\frac{r^2}{w_C^2+w_F^2} \right)$, with $r$ the transverse centre-to-centre distance between the fibre and cavity modes. The cavity mode waist is Gaussian with a $1/e^2$ intensity radius of $w_C=\sqrt{\frac{\lambda}{\pi}\sqrt{LR-L^2}}$, and we take the fibre to also have a Gaussian mode profile with a waist $w_F=2.7\,\mu$m, as specified by the manufacturer (OzOptics Ltd., Ottawa, Canada).\\
 The direct fibre coupling of our cavities leads to minor modifications of the reflected signal compared to an ideal Fabry P\'{e}rot resonator. In the vicinity of an optical resonance, and when all other cavity modes having significant mode overlap with the fibre mode are far-detuned, we consider that light which does not match the cavity mode will be reflected back into the fibre, giving $r_1^a=\sqrt{r_1^2+t_1^2(1-\eta_{FC}^2)}$ in equation (\ref{eqn:contrast}). Conversely we assume that the small fraction of the light which is transmitted from the cavity mode through the fibre mirror, but does not match the fibre mode, is lost into the cladding.\\
The free spectral range $\Delta_{\omega}\simeq c/2L$, where $c$ is the speed of light and $L$ the cavity length, was measured by maintaining one laser at a fixed frequency and tuning a second free-running laser diode from one resonance to the next, and recording the laser frequency at these points. %For the measurement above these frequencies were %$384.76\,$THz and $(381.26\pm 0.01)\,$THz. 
The cavity linewidth was calculated from a double Lorentzian fit to two reflection minima resulting from the two laser frequencies. %They were separated by $235.7\,$GHz in a first and $254.2\,$GHz in a second measurement run with 7 measurements each, while scanning the distance from the fibre block and the cavity chip with an external piezo actuator. 
We further measured the spectral separation $\delta\omega_{HG}$ between the lowest-order resonance and the first and second Hermite-Gaussian modes.  The higher-order modes were made visible by offsetting the fibre core from the cavity mode centre by $r\simeq 1\,\mu$m. The mirror radius of curvature was then calculated using $\delta\omega_{HG}=c(l+m)\arctan (\sqrt{R/L-1})/L$. Here $l$ and $m$ are the mode indices in the $x$ and $y$ directions \cite{saleh}.

 \begin{figure}[!t]\centering
\includegraphics[width=0.6\columnwidth]{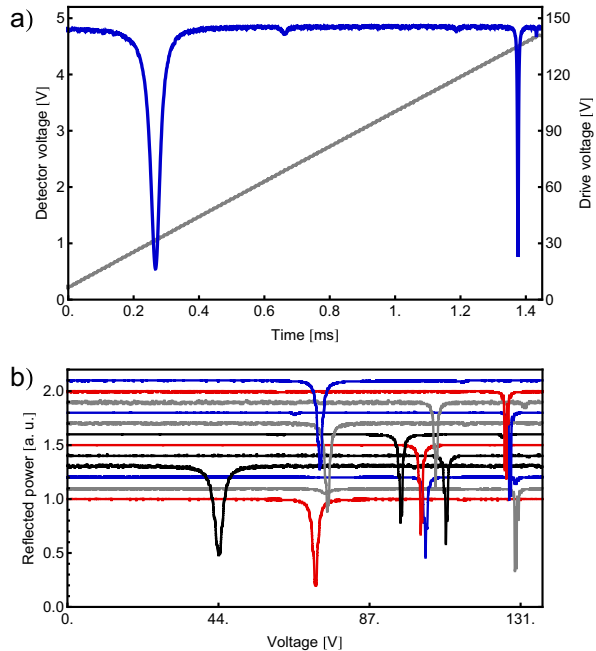}
\caption{\label{fig:opticalPlot} Electrostatic tuning of the optical resonance frequency. a) Reflected power (blue) from an electrostatically tuned microcavity, displaying two resonances of the lowest-order transverse mode. The linear voltage ramp applied to the cantilever is also shown (grey). b) Optical resonances of a row of 12 simultaneously fibre-coupled cavities measured in reflection for a constant distance between the fibre block and the microcavity chip (see text). Each reflection trace is normalised and offset in increments of 0.1 vertical units.}
\end{figure}
Figure \ref{fig:opticalPlot}(a) shows the deflection of a single cantilever micro-mirror through more than one free spectral range. The image shows the reflected power and the applied voltage, and as the voltage is increased linearly (grey line) the cavity traverses two resonances. These are visible as two deep reflection minima, at which the cavity length coincides with two subsequent lowest-order Hermite-Gauss modes of the microcavity \cite{saleh}. The two resonances have different widths due to the non-linear dependence of the cantilever deflection on the applied voltage (see figure \ref{fig:figure2}). Smaller local minima are due to higher-order modes of the optical field. This result indicates that it should be possible to bring an arbitrary number of microcavities into resonance with a desired laser frequency. Indeed, figure \ref{fig:opticalPlot}(b) shows the successful operation of all 12 micro-mirrors on one row of the chip. For this measurement we manually aligned the first and last cantilever in the row to maximise the fibre-cavity mode overlap. In doing so we rely on the fabrication precision of the fibre array and the micro-mirror chip for the precise alignment of all other cavities, and figure \ref{fig:opticalPlot} shows that all twelve reflection traces display a high contrast and weak higher-order mode excitation. We furthermore tilted the fibre block to make the lengths of the first and last cavity in the row match to within one optical linewidth, as visible in the highest and lowest traces in figure \ref{fig:opticalPlot}(b). Subsequently each microcavity was tuned individually, by cantilever deflection, through resonance with a laser which was operated at a constant frequency.  We monitored the distance between the fibre block and micro-mirror chip surface and ensured that the first microcavity in the row was within one resonance full-width at half maximum (FWHM) of its original position, corresponding to $3.7\,$nm, during each measurement.  These results demonstrate the achievement of repeatable coupling, independent tuning and scaled operation.\\
 At the measured cavity length, the microcavities have a calculated $1/$e$^2$ waist radius $w_C=2.16\,\mu$m on the coated fibre tips. The resulting maximum mode overlap to the specified fibre mode waist of $w_F=2.7\,\mu$m is over $95\%$. This is confirmed by the strong suppression of higher-order transverse modes in the reflected signals, and allowed the achievement of the high visibility of $\vartheta=1-P_{min}/P_{max}> 85\,\%$ in the traces in figure \ref{fig:opticalPlot}. A small amount of variation was found in the contrast of the twelve microcavities.  This is due to inherent fluctuations in the fibre core positions which are specified by the manufacturer as $\pm 500\,$nm \textsc{rms}. These quantities indicate that the microcavities will perform well in a scaled cavity QED experiment as the small variability guarantees a high degree of uniformity in performance across the array. 
\begin{figure}\centering
\includegraphics[width=0.5\columnwidth]{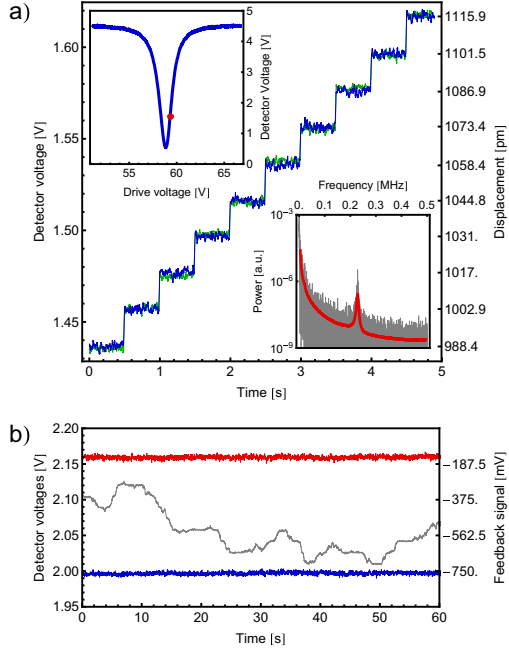}
\caption{\label{fig:stabPic} Stability of operation. a) Step response of a locked cavity around the point of maximum gradient of the optical resonance. The locking setpoint (green) is increased in steps of $20\,$mV and is closely followed by the power reflected from the cavity (blue).  The right scale shows the corresponding increase of the cavity length away from the centre of the resonance. The upper inset shows the reflected power as a function of applied voltage, and a red point marks the steepest point of the curve. The lower inset shows the power spectrum of the detector voltage recorder over one second. The red curve is a running average with a Gaussian window of width $\sigma=5\,$kHz. The mechanical resonance of the cantilever is visible as a small peak at $223\,$kHz. b) Stable operation of two adjacent cavities over one minute.  The blue and red traces show the detector voltages for the two cavities, while the gray line (right scale) is the amplified PID output for one of the cavities.}
\end{figure}
 We have furthermore shown that the length of the microcavities can be stabilised to the order of one picometre. We quantified the length stability of the resonators by locking the reflected power using a proportional-integral-derivative (PID) lock (Stanford Research Systems PID SIM960) with a bandwidth of $100\,$kHz. 
For this experiment, we attached the fibre block to the chip by bonding both to a glass block using UV-curing epoxy (Epotek OG116). The experiments were performed under atmospheric conditions. The results of these measurements, shown in figure \ref{fig:stabPic}, demonstrate that the length of the cavities can be stabilised to variations on the order of picometers.  figure \ref{fig:stabPic}(a) shows the response of a cavity to stepping the set point of the PID lock in steps of $20\,$mV around the point of maximum gradient of the Lorentzian reflection dip (shown in the upper inset). A constant offset between setpoint and detector voltage was removed from the data shown.  The cavity linewidth measured previously gives a fringe full-width at half maximum of $3.7\,$nm, yielding a measured step size on the order of $15\,$pm. The small fluctuations on the voltage steps are on the order of the noise on the set point, and correspond to an average position fluctuation of $\pm 1.1\,$pm. The true position uncertainty is however limited by power fluctuations of $\pm 0.5\%$, due in large part to pointing instability of the laser beam.  These fluctuations increase the uncertainty to $\pm 7.4\,$pm, but can be avoided by employing beam stabilisation or more sophisticated locking methods. The lower inset of the figure shows the noise spectrum of the reflection signal measured for one second at a sampling rate of $1\,$MHz. This measurement shows a weak noise peak at $223\,$kHz, which corresponds to the natural oscillation frequency of this cantilever and is close to the frequency measured by vibrometry for a similar chip (see figure \ref{fig:figure2}). Finally, figure \ref{fig:stabPic}(b) shows the measurement of stable operation over one minute for two adjacent cantilevers. The PID feedback signal for one of the cantilevers, also shown in the image, indicates that the cavity position requires mostly slow correction on a timescale of several seconds. We presume these drifts to be thermal in nature as no effort was made to stabilize the temperature of the setup.\\
 These measurements prove that the devices will allow the stable operation of cavities with far narrower resonances than the current values. Furthermore, intensity stability issues limiting the position accuracy of the current devices can be avoided for smaller linewidths as the implementation of frequency-modulation locking techniques becomes less challenging.
\section{Prospects for implementation}
The microcavity design allows for the inclusion of a variety of quantum emitters. The open access to the optical field in the microcavities and the use of standard components and processing methods guarantee that they can be used to couple photons to atomic systems in combination with atom and ion chip technology.  They can furthermore be implemented with many solid-state emitters as they allow for the inclusion of a transparent crystal layer. The radii of curvature of the micro-mirrors can be increased to several hundreds of micrometres, so that structures to trap and move neutral atoms or ions can be included between the fibre ends and the micro-mirror chip.  Alternatively, part or all of the trapping structures can be fabricated directly on the chip surface in a planar geometry \cite{atomchips, schmidtKaler, nistPlanar, schmidtKalerPlanar,blattPlanar}.\\
 Taking $^{87}$Rb as an exemplary matter qubit, we calculate a vacuum (or half-photon) coupling frequency of $g=\sqrt{3c\lambda^2\gamma/(\pi^2 w_C^2 L_C)}=2\pi\times 0.36\,$GHz on the cycling transition of the D2 line, where $c$ is the speed of light and $\gamma=2\pi\times 3\,$MHz is the amplitude decay rate of the atom's excited state. The resulting single-atom cooperativity, $C_1=g^2/(2\kappa\gamma)=1.48$, compares well with existing cavity QED experiments and makes both single-atom detection and enhanced photon emission possible \cite{rempeNetwork, goldwin}. The current performance can be improved by increasing the mirror reflectivity, which can be achieved using existing smoothing techniques and by replacing the gold coating with a dielectric multilayer \cite{laliotis, biedermann}. It may even be possible to create the reflective layer by nano-structuring of the silicon itself \cite{tunnermannSilicon}. Assuming only the finesse achieved previously in \cite{biedermann} of $6\times 10^4$, the linewidth can be reduced to $2\pi\times 26\,$MHz. This will enable strong coupling with $g\gg \kappa \gg \gamma $ and will increase the single-atom cooperativity to over 800, significantly exceeding the values of currently operated open-access microcavities \cite{rempeNetwork, reichel, meschede}.\\
 Solid-state qubits can also be integrated into the device. One strong candidate for applications in quantum technology is the diamond nitrogen-vacancy (NV) complex, given its long electron and nuclear spin coherence lifetimes and recently demonstrated spin-photon coherence \cite{balasubramanian, togan2010, lukin-nuclear, awschalom, hanson}. Similar properties have also lead to increased interest in many other compatible solid-state emitters \cite{toganDot, SiV, SiC}. NV centers can be created by nitrogen implantation, making the formation of precisely pitched arrays of near-identical emitters possible \cite{toyli,rabeau}. The NV centre's advance into photonic technology is however marred by unfavorable optical properties, such as a low zero-phonon line (ZPL) branching ratio of $\sim 4\,\%$, and this has led to a variety of approaches to include NV centres in optical resonators \cite{hungerNV, diamondPhotonicsReview}. For the cavity dimensions achieved in this work, the waist decreases to $1.95\,\mu$m at the ZPL wavelength of 637nm.  Assuming once again a finesse of $6\times 10^4$ yields a projected single-NV cooperativity of 25. This will enable efficient electronic-state readout and ZPL photon extraction. The creation of a large array of NV centres coupled to the microcavity array will enable the realization of quantum information processing protocols as described in recent work \cite{Nemoto}. The fiber coupling ensures that the scale of such a quantum processor can be increased at will in a modular fashion. The device also brings within reach recently proposed methods for highly sensitive NV-based, cavity-enhanced magnetic gradiometry \cite{Young, Su, budkerCavityMagnetic}.\\
Currently our chips contain 48 independent micro-mirrors, of which on average over $90\%$ function after fabrication.  A further set of processing steps to avoid under-etching the oxide layer under the conductors will allow a drastic reduction of their width, making possible the creation of over 500 cavities on a chip of $1\,$cm$^2$.  The density can be increased even further by replacing the fibre array with integrated photonic waveguides \cite{ladd, mataloni, waltherWaveguide}.
\section{Conclusion}
In summary, we have demonstrated an array of microfabricated, fibre-coupled Fabry-P\'{e}rot microcavities with open access which are independently tunable through more than one free spectral range in the near-infrared.  They present high mechanical resonance frequencies, making them insensitive to environmental acoustic noise. These properties make them ideally suited to form the basis of large-scale CQED systems with ions, atoms or solid-state emitters. The exquisite mechanical properties of single-crystal silicon may furthermore enable applications in precision measurement and cavity optomechanics \cite{taylor, aspel}. This device is therefore a strong candidate for a physical architecture in a scaled quantum technology.\\
 \\
\textbf{acknowledgments} 
This work was funded by the EU FP7 STREP DIAMANT, the EU IP SIQS, the FFG project PLATON NAP, the FWF Wittgenstein prize and the TUW project DDQuT. We thank M. St\"{u}we for his participation in the fabrication of the micromirror chip, M. Zach and H. Hartmann for workshop assistance, J. Leitner and S. Wald for their assistance in the measurements and F. Keplinger for providing access to the white-light interferometer.

\end{document}